\documentclass[%
 article,
 twocolumn,
nofootinbib,
 amsmath,amssymb,
 aps,
]{revtex4-1}

\usepackage{graphicx}
\usepackage{dcolumn}
\usepackage{bm}
\usepackage{hyperref}
\usepackage{float}%

\begin{document}


\title{First-Principles Study of Strain Effect on Thermoelectric Properties of LaP and LaAs}

\author{Chia-Min Lin}
\affiliation{Department of Physics, University of Alabama at Birmingham, Birmingham, Alabama 35294, USA}

\author{Wei-Chih Chen}
\affiliation{Department of Physics, University of Alabama at Birmingham, Birmingham, Alabama 35294, USA}

\author{Cheng-Chien Chen$^*$}
\affiliation{Department of Physics, University of Alabama at Birmingham, Birmingham, Alabama 35294, USA}
\footnotetext{Corresponding author email: chencc@uab.edu}

\begin{abstract}
Rare-earth monopnictides have attracted much attention due to their unusual electronic and topological properties for potential device applications.
Here, we study rock-salt structured lanthanum monopnictides LaX (X = P, As) by density functional theory (DFT) simulations.
We show systematically that a meta-GGA functional combined with scissor correction can efficiently and accurately compute electronic structures on a fine DFT $k$-grid, which is necessary for converging thermoelectric calculations.
We also show that strain engineering can effectively improve thermoelectric performance.
Under the optimal condition of 2\% tensile strain and carrier concentration $n=3\times10^{20}~\textrm{cm}^{-3}$, LaP at temperature 1200 K can achieve a figure of merit $ZT$ value $>2$, which is enhanced by 90\% compared to the unstrained value.
With carrier doping and strain engineering, lanthanum monopnictides thereby could be promising high-temperature thermoelectric materials.
\end{abstract}

\pacs{Valid PACS appear here}


\maketitle


\section{INTRODUCATION}
Thermoelectric materials can directly convert heat into electricity, and they have various existing and potential applications in energy sector and powering industry~\cite{yang2006review,gaultois2013data,zhang2015review,petsagkourakis2018thermoelectric,wolf2019high,beretta2019thermoelectrics}.
Thermoelectric technology is eco-friendly, where the thermal source can be waste heat, fuels, as well as geothermal and solar energy.
Small-scale thermoelectric devices can be made as power supply for electricity generation or as refrigerator for cooling purpose.
In all these applications, achieving a high thermoelectric efficiency is crucial.
The performance of thermoelectric materials can be determined by the dimensionless figure of merit $ZT$:
\begin{equation}\label{eq:ZT}
 ZT=\frac{S^2\sigma T}{\kappa} = \frac{S^2\sigma T}{(\kappa_e + \kappa_L)} .
\end{equation}
Here, the Seebeck coefficient $S$ is a measure of induced voltage due to temperature difference across a material.
$\sigma$ is the electrical conductivity, $T$ is the temperature, and $\kappa$ is total thermal conductivity,
which has both electronic ($\kappa_{e}$) and lattice ($\kappa_{L}$) contributions.
$ZT\geq 1$ is considered suitable for practical thermoelectric applications, 
and there is an ongoing need to develop larger $ZT$ materials that are functional at different temperatures~\cite{dehkordi2015thermoelectric, twaha2016comprehensive}.

Several approaches have been proposed to improve thermoelectric performance, ranging from band- and nano-structure engineering~\cite{heremans2008enhancement, pei2011convergence, hinsche2012thermoelectric,xi2016band, Isaacs2019}, application of strain~\cite{kaur2017strain,wei2020strain}, to the search of new quantum materials~\cite{xu2017review, gooth2018review}.
Based on the definition in Eq. (\ref{eq:ZT}), a high $ZT$ value requires large $S$, large $\sigma$, and/or small $\kappa$.
In many cases, however, these requirements cannot be satisfied simultaneously due to the trade-off relationships between different parameters.
For example, a larger carrier concentration can enhance $\sigma$, but it can reduce $S$ as well.
Also, materials with high $\sigma$ often exhibits large $\kappa_e$.
Moreover, while $ZT$ has an explicit linear $T$ dependence, $S$ can degrade at high temperature due to bipolar excitation of both hole and electron carriers~\cite{glassbrenner1964thermal,berman1978thermal,shi2015connecting,gong2016investigation}.
Therefore, increasing $ZT$ remains a nontrivial and challenging task.

In this paper, we study the effects of doping and strain on lanthanum monopnictides LaX (X = P, As), which are promising high-temperature thermoelectric materials.
LaX with the rock-salt structure belong to a large group of rare-earth pnictides \cite{hulliger1979rare}, and they have been widely investigated due to their thermal stability, mechanical strength, as well as exotic electronic and topological properties~\cite{kimura1995optical, petukhov1996electronic, vaitheeswaran2002electronic, shirotani2003pressure, pagare2005high, gupta2010first, kumar2016observation, zhou2018LaAs, zhou2019thermoelectric}. 
However, the rare-earth element and its correlated orbitals near the Fermi level have posed challenges for first-principles band-theory studies.
In particular, it has been a debate whether the systems are metallic or semiconducting~\cite{hasegawa1980electronic,norman1984electronic,gokouglu2008lattice,ciftci2008first}. 
Deligoz {\it et al.} found that both LaP and LaAs are metals, based on density functional theory (DFT) with the local-density approximation (LDA)~\cite{deligoz2007electronic}.
However, Charifi {\it et al.} found that LaN, LaP, and LaAs are semiconductors using both LDA and the generalized gradient approximation (GGA) calculations~\cite{charifi2008phase}; Shoaib {\it et al.} \cite{shoaib2013structural} also reached a similar conclusion by using the Wu and Cohen GGA functional~\cite{wu2006more}.
Moreover, Yan {\it et al.}~\cite{yan2014theoretical} and Khalid {\it et al.}~\cite{khalid2020hybrid} reported non-zero energy gaps in LaP and LaAs, by using a more sophisticated Heyde-Scuseriae-Ernserh hybrid functional (HSE06).
Since the band structure and energy gap play crucial roles in determining transport and thermoelectric properties, it is important to ensure accurate description and computation of the electronic structures.

\begin{figure*}
\includegraphics[width=1.0\linewidth]{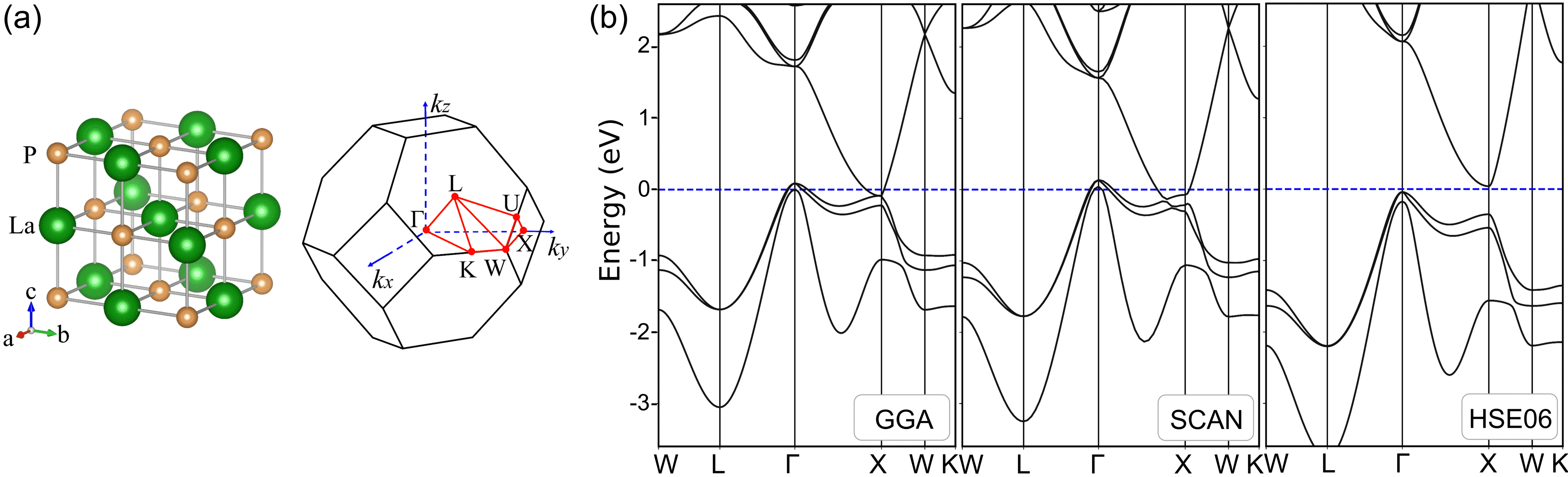}
\caption{
(a) LaP face-centered cubic structure and its Brillouin zone.
(b) Band structures along high-symmetry paths of the Brillouin zone computed respectively by GGA, SCAN, and HSE06 functionals for unstrained LaP.
The zero energy corresponds to the Fermi level.
}
\label{Fig.1}
\end{figure*}
  
In the following, we perform systematic calculations by considering three different rungs of the ``Jacob's ladder"~\cite{perdew2001jacob}, including GGA, meta-GGA, and hybrid DFT functionals.
We show that meta-GGA calculations combined with a scissor correction method can provide accurate and converged thermoelectric results on a refined $k$-grid, which otherwise is not possible with a computationally much more expensive hybrid functional. We also show that isotropic strain and carrier doping are effective means to manipulate the transport properties of LaX for potential high-temperature thermoelectric applications.

\section{COMPUTATIONAL METHODS}
The properties of LaX (X = P, As) are investigated by density functional theory (DFT) \cite{hohenberg1964inhomogeneous,kohn1965self} using a plane wave pseudopotential method as implement in the Vienna {\it ab initio} simulation package (VASP)~\cite{kresse1996efficiency,kresse1996efficient}.
We use a plane wave energy cutoff of 500 eV, which is 30\% larger than the recommended value in the VASP pseudopotential files, and it suffices to converge the DFT total energy with a difference  $< 10^{-4}$ eV/atom.
For each material, we first relax the lattice parameters to obtain the unstrained rock-salt structures using the Perdew-Burke-Ernzerhof generalized gradient approximation (GGA-PBE) functional~\cite{perdew1996generalized}.
After structure relaxation, we then perform self-consistent electronic calculations with spin-orbit coupling respectively for three functionals: the GGA-PBE functional, the meta-GGA strongly-constrained and appropriately-normed (SCAN) functional~\cite{sun2015strongly}, and the hybrid Heyde-Scuseriae-Ernserh (HSE06) functional~\cite{heyd2003hybrid,brothers2008accurate}.
Studying different advanced functionals beyond GGA will help determine the electronic band gap more accurately.
The convergence criteria of self-consistent and structure calculations are set to 10$^{-5}$ eV/unit cell and 10$^{-4}$ eV/$\text{\normalfont\AA}$, respectively. The Monkhorst-Pack sampling scheme~\cite{monkhorst1976special} is used with a $\Gamma$-centered $k$-point mesh with a grid size of $11\times 11 \times 11$ (resolution = 0.01$\times$2$\pi$/$\text{\normalfont\AA}$) points over the Brillouin zone.

For transport quantities -- the Seebeck coefficient $S$, electrical conductivity $\sigma$, and electronic thermal conductivity $\kappa_e$ -- a much refined $k$-grid is needed to converge the calculations (see the Supplemental Material~\cite{SM} for convergence tests). Therefore, we compute transport properties using the meta-GGA SCAN functional with spin-orbit coupling on a denser $k$-gird of $41 \times 41 \times 41$ (resolution = 0.006$\times$2$\pi$/$\text{\normalfont\AA}$) points. HSE06 calculations on such a $k$-grid is computationally too expensive to perform. On top of the meta-GGA calculations, we further introduce a scissor operator to correct the band gap, which plays a crucial role in determining the thermoelectric performance.
With inputs from VASP calculations, transport quantities are computed by the \textsc{BoltzTraP2} package~\cite{madsen2018boltztrap2}.
Based on a linearized version of the Boltzmann transport equation, \textsc{BoltzTraP2} evaluates the transport distribution function~\cite{mahan1996best} under the rigid-band and constant relaxation time approximations.
The transport coefficients $S$, $\sigma$, and $\kappa_e$ can be computed directly once the transport distribution function is determined.

For phonon spectra and lattice thermal conductivity $\kappa_{L}$, we utilize respectively the \textsc{Phonopy} package~\cite{phonopy} and the \textsc{Phono3py} code~\cite{phono3py,phono3py-app}. \textsc{Phonopy} computes the phonon dispersion based on the harmonic approximation, and it enables us to assure that the rock-salt structures of LaX (X = P, As) under study are dynamically stable (i.e. with only positive phonon modes). 
\textsc{Phono3py} solves linearized phonon Boltzmann transport equation with single-mode relaxation time approximation.
Finite atomic displacements and supercell approaches with respectively $5\times 5 \times 5$ and $4 \times 4 \times4$ supercells are adopted to compute the second- and third-order force constants, which are required for evaluating phonon lifetimes and $\kappa_L$.
A phonon $q$-point sampling mesh of $21 \times 21 \times 21$ points and a real-space cutoff approach up to the fourth neighbor ($N$-cutoff = 4, with a cutoff radius $\sim 6.1~\text{\normalfont\AA}$) are utilized to reduce the computational demand for the third-order force constants. 
The corresponding VASP calculations are based on the GGA-PBE functional without the spin-orbit coupling.
Further \textsc{Phonopy} and \textsc{Phono3py} results and convergence tests are given in the Supplemental Material~\cite{SM}.

Finally, the theoretical crystal structure in our study is visualized by the VESTA software~\cite{momma2011vesta}.
  
\begin{figure*}
\includegraphics[width=1.0\linewidth]{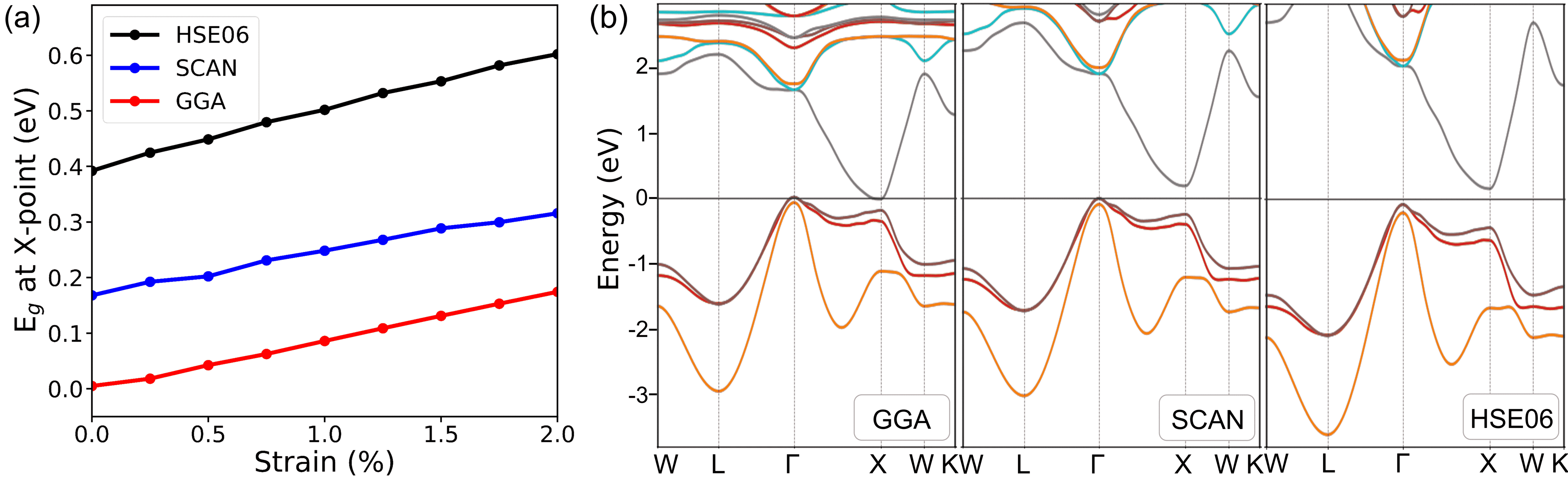}
\caption{
(a) Band gaps at the X point for LaP computed as a function of strain using different functionals.
(b) Band structures computed respectively by GGA, SCAN, and HSE06 functionals for LaP under 2\% tensile strain.
The zero energy corresponds to the Fermi level.
}
\label{Fig.2}
\end{figure*}

\section{RESULTS AND DISCUSSION}
We begin by discussing LaP, which has a stable face-centered cubic structure with space group symmetry $Fm\bar{3}m$ (No. 225). This rock-salt structure and its first Brillouin zone are shown in Fig. 1(a).
In our structure relaxation calculation, the equilibrium lattice parameter of LaP is found to be $a=4.278~\text{\normalfont\AA}$.
With the unstrained structure, we then compute the electronic band structures including spin-orbit coupling respectively for three different DFT functionals: GGA, meta-GGA SCAN, and HSE06. 
As shown in Fig. 1(b), the GGA (left panel) and meta-GGA SCAN (middle panel) results indicate that LaP is a semimetal with band crossing (between predominantly La $d$- and P $p$-orbitals) near the X point. However, the HSE06 result (right panel) shows a band gap opening $\sim0.39$ eV at the X point, indicating that LaP is a semiconductor, which agrees with previous studies using advanced DFT functionals~\cite{shoaib2013structural,yan2014theoretical, zhou2019thermoelectric}.
It is known that GGA typically underestimates the energy gap. While the meta-GGA SCAN functional can improve the band gap estimation~\cite{patra2019efficient}, this improvement often remains insufficient.
Basically, the simple functional forms of GGA and meta-GGA are not flexible enough to accurately reproduce both exchange correlation energy and its charge derivative~\cite{perdew2001jacob}.
On the other hand, the advanced hybrid functional HSE06 provides a more sophisticated nonlocal expression and leads to a more precise gap prediction.

Figure 2(a) shows how the X point band gap evolves with strain for different functionals. The gap values are obtained by band structure interpolations using \textsc{BoltzTraP2} with inputs from VASP self-consistent calculations.
Overall, the energy gap would increase under tensile strain, and its rate of increase with strain is similar for all three functionals.
Figure 2(b) shows the interpolated band structures at the maximum value of $+2\%$ (tensile) strain under study, where the X point band gaps are 0.18 eV, 0.32 eV, and 0.60 eV, respectively for GGA, SCAN, and HSE06.
We note again that converging the DFT total energy is much easier than converging thermoelectric calculations.
As shown in the systematic studies in the Supplemental Material~\cite{SM}, a very refined $k$-grid is needed to satisfactorily converge transport properties.
For the results presented below, we adopt a dense $k$-grid of $41 \times 41 \times 41$ points, so the calculations can be performed only with the GGA and meta-GGA SCAN functionals.
In addition, we further employ a scissor method and correct the GGA and meta-GGA band gaps to that of HSE06 based on Fig. 2(a), which provides a roadmap for band gap correction as a function of strain.

Figures 3(a) and 3(b) show the Seebeck coefficients $S$ of unstrained LaP as a function of chemical potential $\mu$ at various temperatures, using respectively the GGA and meta-GGA SCAN functionals.
The red vertical dashed line in Fig. 3 indicates the Fermi level $E_f$.
As expected, $S$ is positive when hole carriers dominate ($\mu - E_f < 0$), and $S$ is negative when electron carriers dominate ($\mu - E_f  > 0$).
Compared to the GGA result [Fig. 3(a)], the peak strength of $S$ in meta-GGA [Fig. 3(b)] is slightly enhanced due to an enlarged band gap.
Figures 3(c) and 3(d) show similar calculations respectively for GGA and SCAN functionals with scissor correction. In both cases, the conduction bands overall are shifted up in energy by matching the X point gap to that of HSE06.
The results of GGA with scissor correction [Fig. 3(c)] resemble closely those of SCAN with scissor correction [Fig. 3(d)],
both showing that $S$ is further enhanced after the gap correction.
Therefore, the above results show that obtaining a proper band gap can play a dominant role in determining thermoelectric behaviors.

\begin{figure}
\includegraphics[width=1.0\linewidth]{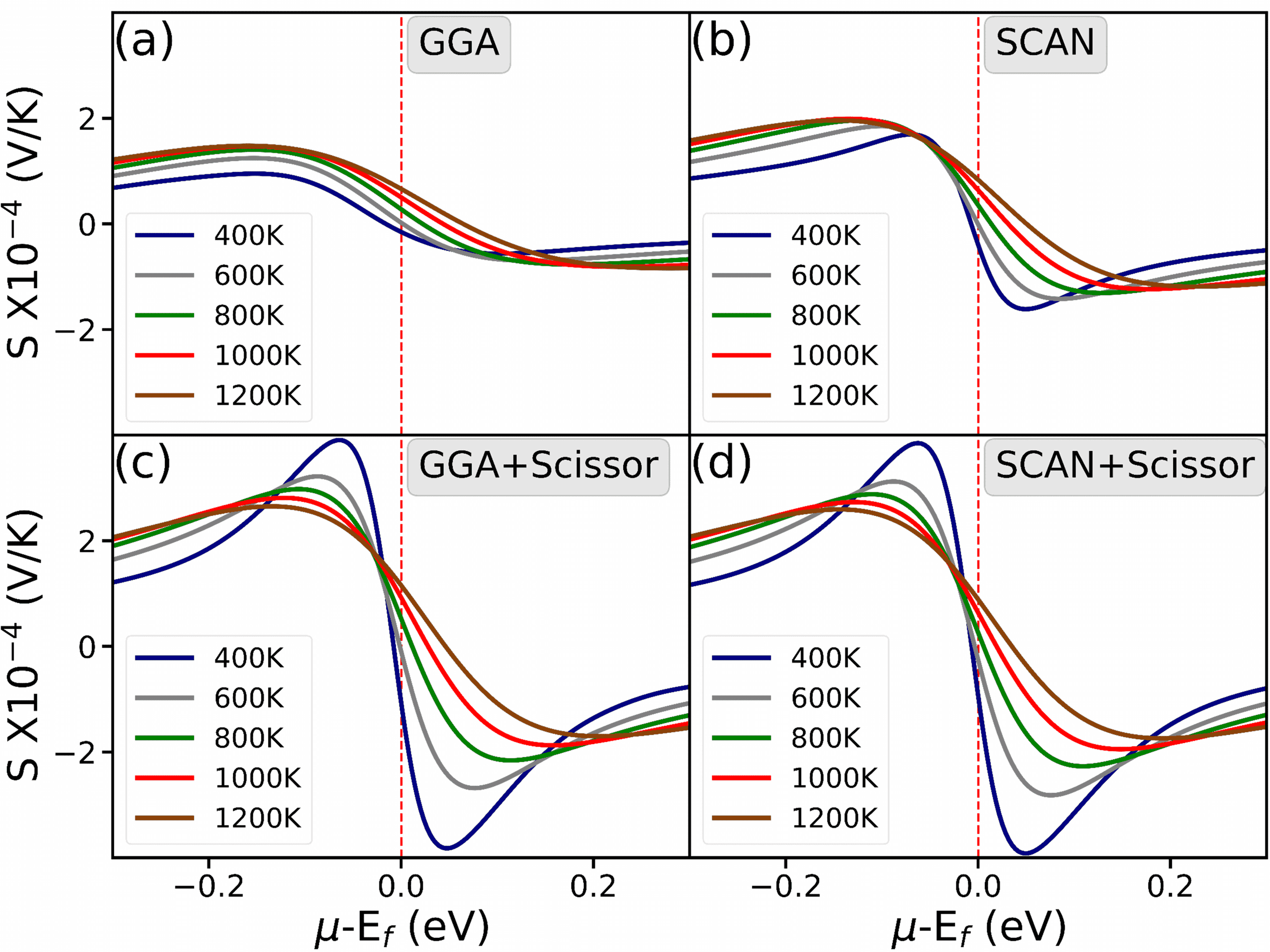}
\caption{
Seebeck coefficients as a function of the chemical potential $\mu$ at different temperatures, computed for unstrained LaP using respectively (a) GGA functional,(b) SCAN functional, (c) GGA functional with scissor correction, and (d) SCAN functional with scissor correction.
}
\label{Fig.3}
\end{figure}

\begin{figure}
\includegraphics[width=1.0\linewidth]{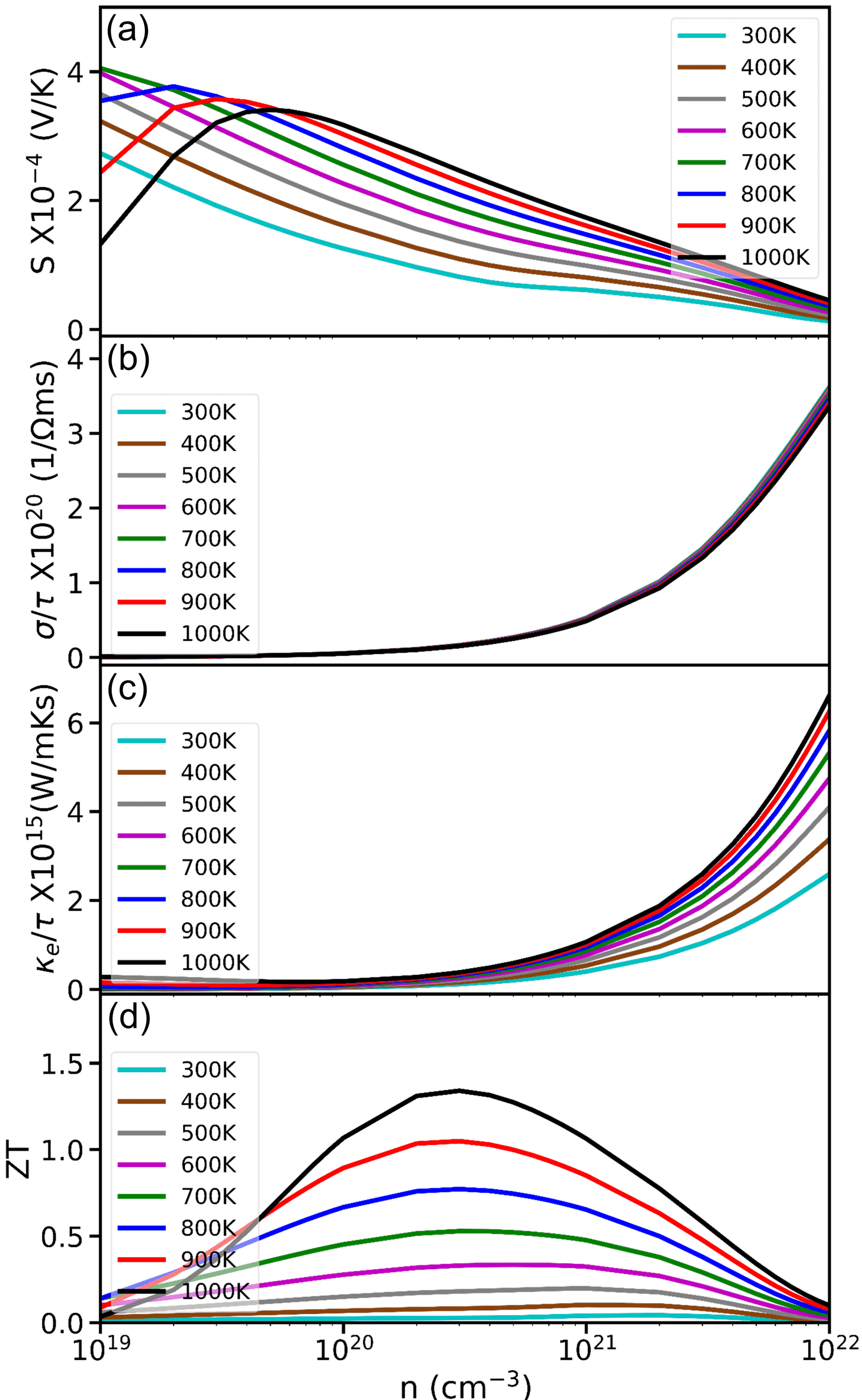}
\caption{
Thermoelectric properties of unstrained LaP: (a) Seebeck coefficient $S$, (b) electrical conductivity $\sigma$ divided by the relaxation time $\tau$, (c) electronic thermal conductivity $\kappa_{e}$ divided by $\tau$, and (d) $ZT = S^{2}\sigma T/\kappa$, as a function of carrier concentration $n$ in the temperature range $T = 300 - 1000$ K. The calculations are performed using the SCAN functional with scissor correction. The horizontal axis is in a log scale.
}
\label{Fig.4}
\end{figure}

Figure 4(a) shows \textsc{BoltzTraP2} calculations of the Seebeck coefficient $S$ as a function of carrier density $n$ at various temperature $T$, using the SCAN functional with scissor correction for unstrained LaP.
Overall, $S$ is enhanced with increasing $T$ or decreasing $n$. These behaviors can be understood qualitatively using a nearly free electron picture with parabolic band and energy-independent scattering approximations~\cite{PhysRev.133.A1143, NatMat2008}:
\begin{equation}\label{eq:freeS}
S = \frac{8\pi^2 k^2_B}{3eh^2} m^* T (\frac{\pi}{3n})^{2/3}.
\end{equation}
Here, $e$, $k_B$, $h$, and $m^*$ are respectively electron charge, Boltzmann constant, Plank constant, and carrier effective mass. Equation (\ref{eq:freeS}) predicts that $S$ depends linearly on $T$ and varies as $n^{-2/3}$, which is consistent with Fig. 4(a) in the high carrier concentration regime. 
On the other hand, $S$ also can decrease with increasing temperature in the low carrier concentration region, as shown in Fig. 4(a). This unusual situation is caused by a bipolar conduction (or finite-temperature excitation of both hole and electron carriers)~\cite{glassbrenner1964thermal,berman1978thermal,shi2015connecting,gong2016investigation}.
The bipolar effect is more prominent in narrow-gap semiconductors ($\lesssim$ 0.5 eV), and it can lead to degrading thermoelectric performance.

Figure 4(b) shows the electrical conductivity $\sigma$ divided by the relaxation time $\tau$. Since it is more difficult to compute $\tau$ from first principles due to the complexity of different scattering mechanisms, here we simply adopt a typical value of $\tau = 1 \times10^{-14}$ s.
In general, $\sigma$ shows a weak temperature dependence and decreases slightly with increasing $T$. Also as expected, $\sigma$ increases with increasing $n$ ($\sigma \propto n$), which is opposite to the behavior of Seebeck coefficient ($S \propto n^{-2/3}$). In addition, $\sigma$ is expected to be inversely proportional to the effective mass $m^*$, while $S$ depends linearly on $m^*$. For the above reasons, it is thereby challenging to enhance a material's thermoelectric power factor ($\equiv S^2 \sigma$).

Figure 4(c) shows the electronic thermal conductivity $\kappa_{e}$ divided by $\tau$. In general, $\kappa_e$ has a similar dependence on $n$ as $\sigma$. However, $\kappa_e$ has a stronger temperature dependence and increases with increasing $T$.
These behaviors obey qualitatively the Wiedemann-Franz law~\cite{chester1961law}:
\begin{equation}\label{eq:WFlaw}
\kappa_{e}=\frac{\pi^{2}}{3} (\frac{k_{B}}{e})^{2}\sigma T \equiv L \sigma T,
\end{equation}
which is based on the fact that both heat and electrical transport involve free charge carriers in metals.
Here, $L = 2.44\times 10^{-8}~W\Omega K^{-2}$ is the free-electron Lorentz number.
Since $ZT \equiv S^2\sigma T/(\kappa_e + \kappa_L)$, for materials with high $\kappa_e$ and low $\kappa_L$, or when $\kappa_e \gg \kappa_L$ at very high temperature, Eq. (\ref{eq:WFlaw}) dictates that $ZT \simeq S^2/L$.
In this case, a less dispersive flat band (which results in larger effective $m^*$ and $S$) is expected to show a higher $ZT$ value. 

The actual $ZT$ value of unstrained LaP is shown in Fig. 4(d), which is obtained from $S$, $\sigma$, and $\kappa_e$ in Figs. 4(a)-4(c), as well as from the lattice thermal conductivity $\kappa_L$ (to be discussed shortly in Fig. 5).
The $ZT$ value is highly dependent on the carrier concentration $n$ and temperature $T$. 
For example, $ZT$ has a peak value of 0.77 at $T= 800$ K near $n=3\times10^{20}~\textrm{cm}^{-3}$.
The maximum $ZT$ value is enhanced to 1.34 at $T=1000$ K near the same $n$.
Therefore, in our later discussion of strain effect on LaP, we fix the carrier concentration to be $n=3\times10^{20}~\textrm{cm}^{-3}$ for potential optimal thermoelectric performance.

\begin{figure}
\includegraphics[width=1.0\linewidth]{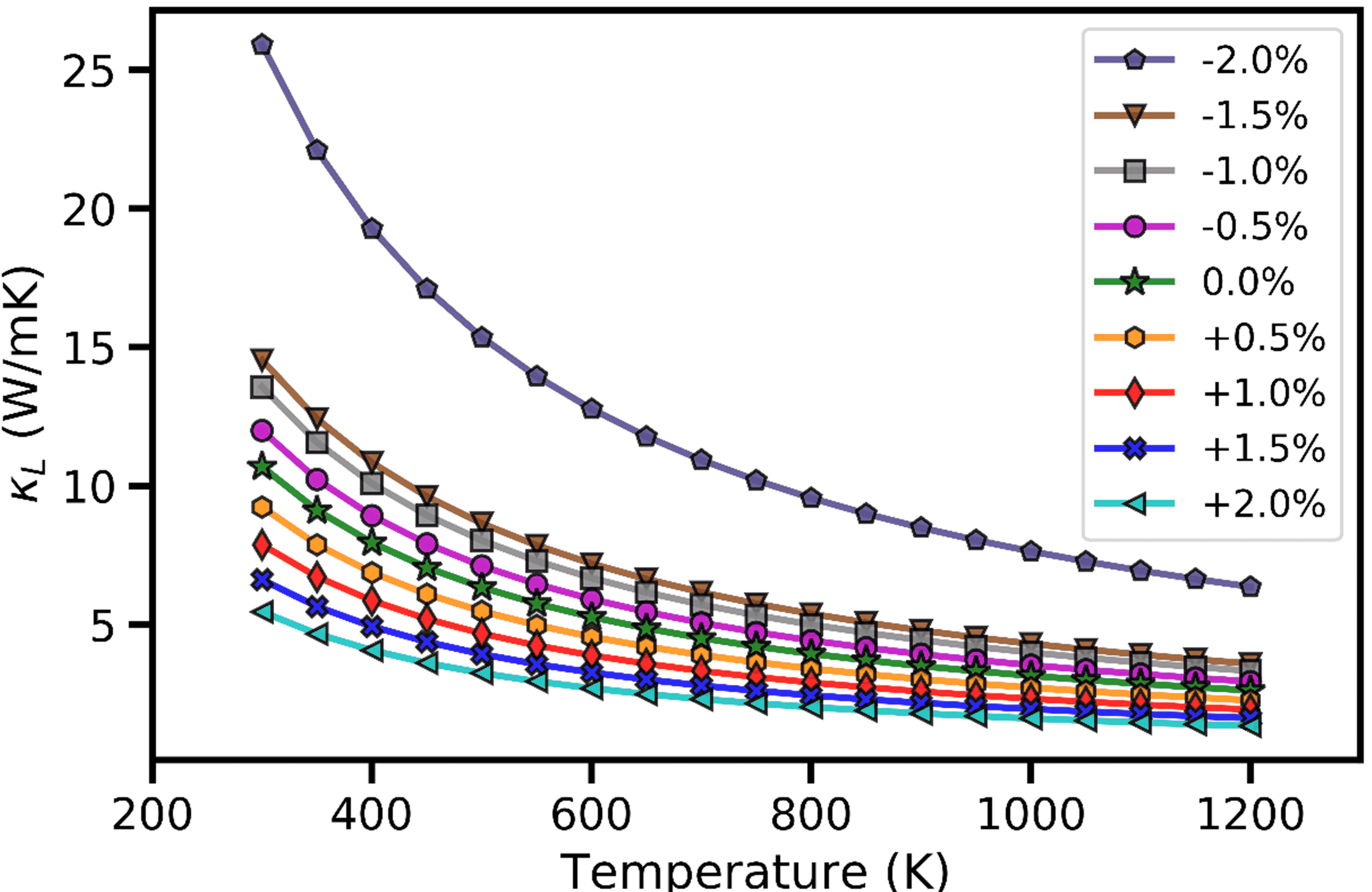}
\caption{
Lattice thermal conductivity $\kappa_{L}$ as a function of temperature for LaP in the strain range $-2\%$ to $+2\%$.
}
\label{Fig.5}
\end{figure}

Figure 5 shows the lattice thermal conductivity $\kappa_{L}$ of LaP as a function of $T$ in the -2\% (compressive) to +2\% (tensile) strain range. 
Notably, $\kappa_L$ decreases with increasing $T$, and behaves as $\kappa_L \sim T^{-1}$ at high temperature~\cite{toberer2011phonon}.
This behavior can be understood qualitatively using elementary kinetic theory and the Debye model:
\begin{equation}
\kappa_L = \frac{1}{3} v^2_s c_v \tau_s.
\end{equation}
Here, $v_s$ and $c_v$ are respectively the phonon velocity and specific heat. In Debye model, $v_s$ has no temperature dependence, and $c_v$ is only weakly dependent on $T$ above the Debye temperature $\Theta_D$. $\tau_s$ is the phonon relaxation time (or equivalently, $\tau^{-1}_s$ is the phonon scattering rate). Therefore, the temperature dependence of $\kappa_L$ at $T\gg \Theta_D$ is governed by $\tau_s$, which is inversely proportional to the phonon occupation number $n_s(\mathbf{q}) \sim k_B T/\hbar\omega_s(\mathbf{q})$ at high temperature. Therefore, $\kappa_L$ scales as $\sim T^{-1}$ at $T \gg \Theta_D$.
Also as shown in Fig. 5, $\kappa_L$ is reduced (enhanced) by a tensile (compressive) strain, due to a softening (hardening) of the phonon velocity.
Finally, we note that \textsc{Phono3py} considers only three-phonon scattering processes.
If four-phonon processes are also included, $\kappa_L \sim (T + \alpha T^2)^{-1}$ will be further reduced (and $ZT$ will be further enhanced accordingly). 

\begin{figure}
\includegraphics[width=1.0\linewidth]{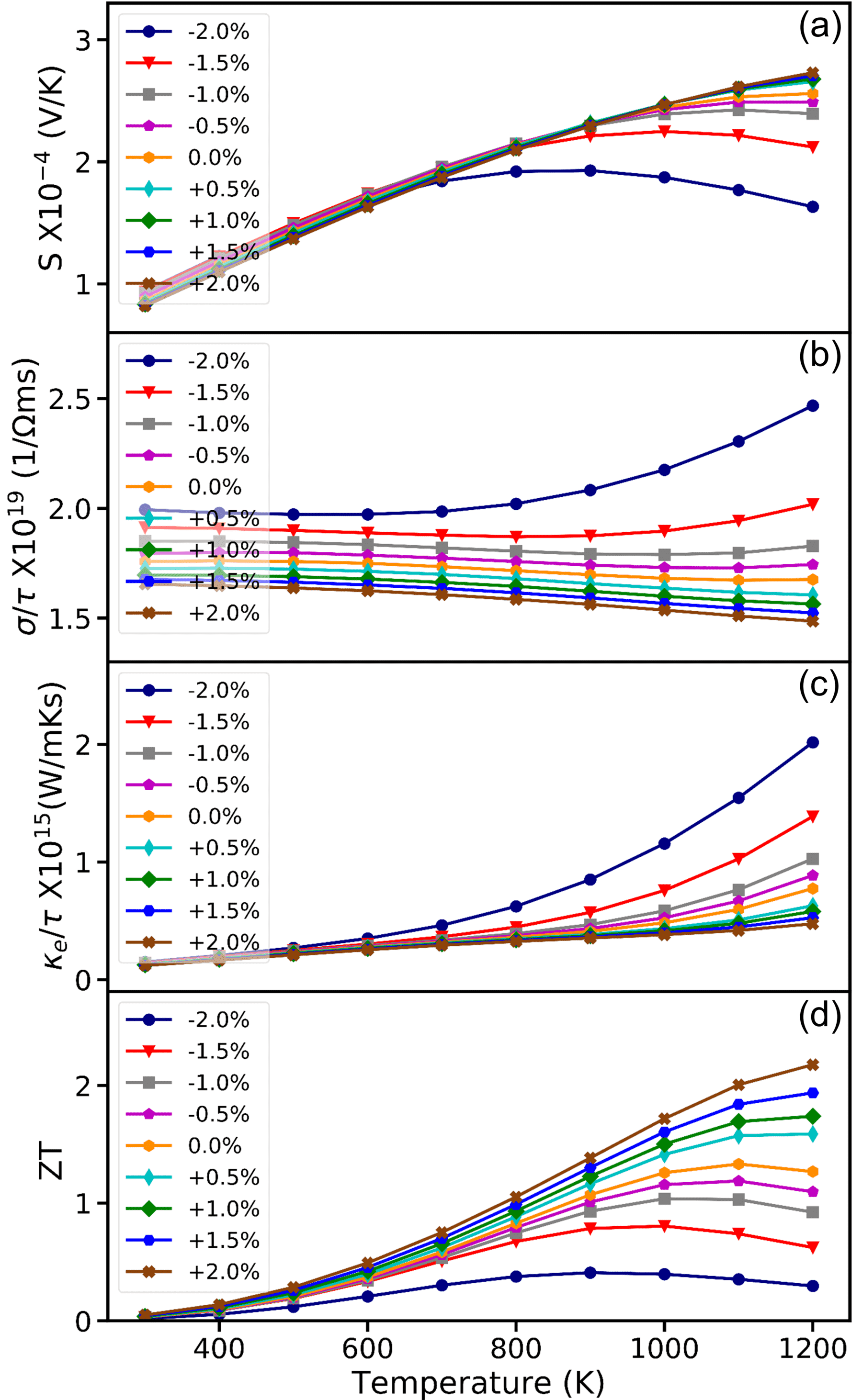}
\caption{
Thermoelectric properties of LaP in the strain range $-2\%$ to $+2\%$: (a) Seebeck coefficient $S$, (b) electrical conductivity $\sigma$ divided by the relaxation time $\tau$, (c) electronic thermal conductivity $\kappa_{e}$ divided by $\tau$, and (d) $ZT = S^{2}\sigma T/\kappa$, as a function of temperature. The calculations are performed using the SCAN functional with scissor correction at carrier concentration $n=3\times10^{20}~\textrm{cm}^{-3}$.
}
\label{Fig.6}
\end{figure}
  
We next discuss the effect of strain on other transport coefficients and the $ZT$ value, by fixing the carrier concentration at $n=3\times10^{20}~\textrm{cm}^{-3}$.
Figure 6(a) shows the temperature evolution of Seebeck coefficient $S$ for strain values ranging from $-2\%$ to $+2\%$.
As discussed before, $S$ will increase with $T$, except when the bipolar effect sets in at higher temperature. In addition, $S$ can be enhanced by a tensile strain, mainly due to an increased carrier effective mass $m^*$ by strain engineering.
This is consistent with the behavior of the electrical conductivity $\sigma$ shown in Fig. 6(b), where $\sigma$ is reduced by a tensile strain (as $\sigma \propto 1/m^*$).
Similar to $\sigma$, the electronic thermal conductivity $\kappa_e$ is reduced by a tensile strain, while $\kappa_e$ uplifts with increasing $T$. These behaviors are consistent with the Wiedemann-Franz law in Eq. (\ref{eq:WFlaw}).
Since a tensile strain can enhance $S$ and meanwhile reduce both $\kappa_e$ and $\kappa_L$, the $ZT$ value is expected to increase accordingly. Indeed, Fig. 6(d) shows that the $ZT$ value strongly depends on the temperature and applied strain.
In general, $ZT$ is enhanced with increasing $T$ and positive (tensile) strain.
At $T=1200$ K, the $ZT$ value of LaP under a $+2\%$ tensile strain can exceed 2, which is enhanced by 90\% compared to the unstrained value.
  
\begin{figure}
\includegraphics[width=1.0\linewidth]{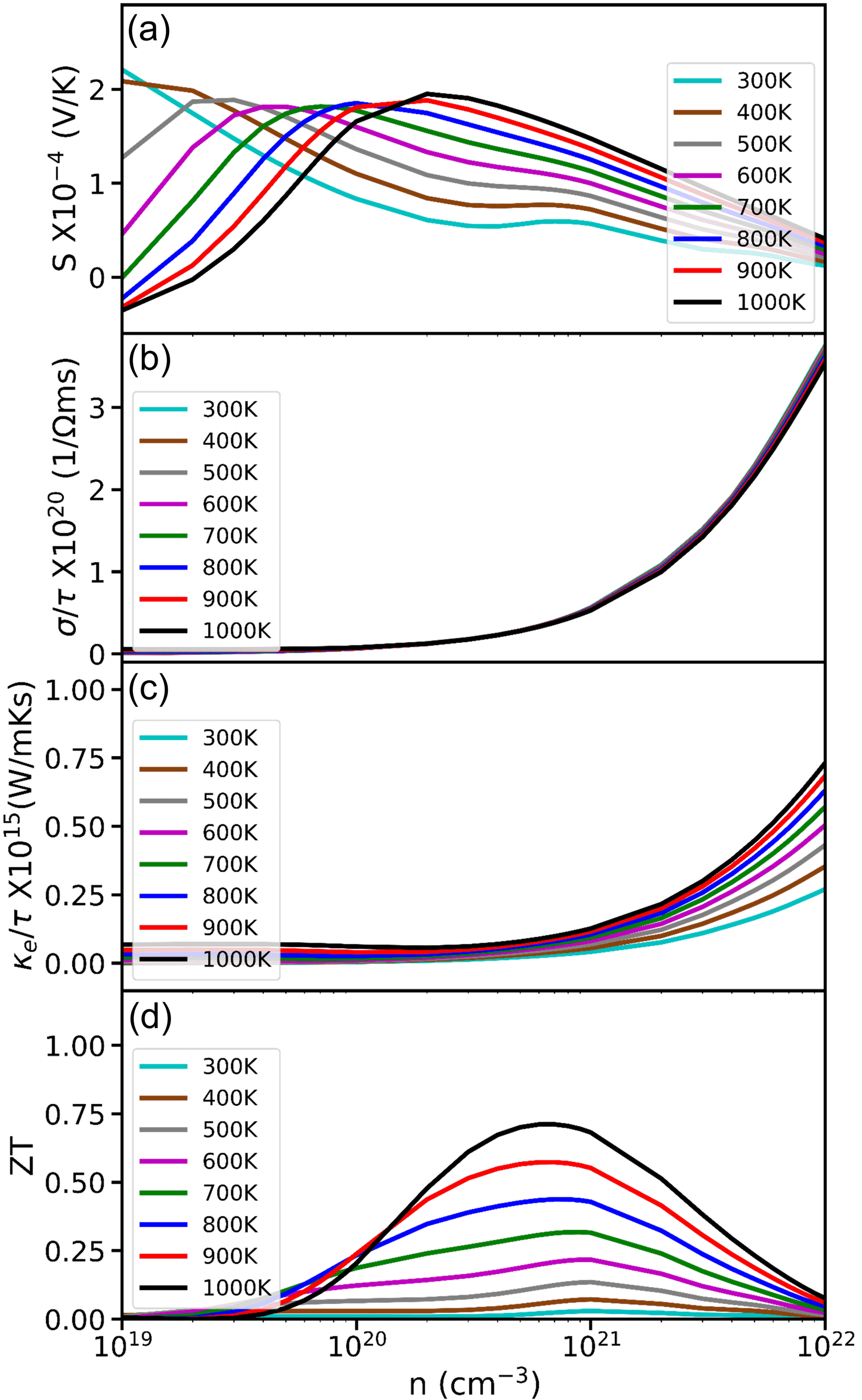}
\caption{
Thermoelectric properties of unstrained LaAs: (a) Seebeck coefficient $S$, (b) electrical conductivity $\sigma$ divided by the relaxation time $\tau$, (c) electronic thermal conductivity $\kappa_{e}$ divided by $\tau$, and (d) $ZT = S^{2}\sigma T/\kappa$, as a function of carrier concentration $n$ in the temperature range $T = 300 - 1000$ K. The calculations are performed using the SCAN functional with scissor correction. The horizontal axis is in a log scale.
}
\label{Fig.7}
\end{figure}

We last turn our discussion to LaAs, which is also dynamically stable in the rock-salt structure like LaP~\cite{SM}.
The relaxed lattice parameter of unstrained LaAs is found to be $a=4.379~\text{\normalfont\AA}$.
LaAs has an overall band structure resembling that of LaP, albeit with a smaller energy gap at the X point~\cite{khalid2018topological}. This reduced gap size is consistent with electronegativity consideration: The band gap of a binary A-B compound is positively correlated with the atoms' electronegativity difference $| \chi_\textrm{A} - \chi_\textrm{B}|$~\cite{chi_gap}.
In Paulin scale, $\chi_\textrm{P} > \chi_\textrm{As} > \chi_\textrm{Sb} > \chi_\textrm{Bi} > \chi_\textrm{La} = 1.1$. Therefore, the band gap of LaX would decease when X moves down along the pnictogen group in the periodic table, which agrees with DFT calculations~\cite{khalid2018topological}.

Figure 7(a) shows the Seebeck coefficient $S$ of LaAs as a function of carrier concentration $n$ at various temperatures $T$. Compared to LaP, there are two salient features. First, the bipolar effect on LaAs is more severe, due to a narrower band gap which enhances thermal excitations of both charge and hole carriers.
Second, LaAs has a smaller Seebeck coefficient. For example, the peak value of $S=1.97$ at $T=1000$ K in LaAs is reduced to roughly 60\% of that in LaP. Because of a larger $p$-orbital spread, LaAs has a more dispersive band structure (with a smaller effective mass $m^*$), so a smaller $S$ is anticipated.

Figures 7(b) and 7(c) show respectively the electrical conductivity $\sigma$ and the electronic thermal conductivity $\kappa_e$, divided by the relaxation time $\tau$. Their overall dependences on the carrier concentration and temperature follow similar trends to those discussed in Figs. 4(b) and 4(c) for LaP. The resulting $ZT$ value of LaAs is shown in Fig. 7(d). At $T=800$ K, $ZT$ is peaked at $n=7 \times 10 ^{20}~\textrm{cm}^{-3}$; the peak value of $ZT=0.44$ is $\sim 60\%$ that of LaP at the same temperature. Therefore, LaAs has a lower thermoelectric performance. Nevertheless, it is expected that the maximum $ZT$ value could be boosted to near unity at $T\ge 1000$ K with applied strain.
     
Finally, we note that LaSb and LaBi have even smaller band gaps and more dispersive bands (with smaller $m^*$). On general grounds, we thereby do not expect LaSb and LaBi to be of practical thermoelectric applications. However, these materials can exhibit nontrivial topological properties~\cite{LaX_TI_Lin, LaBi_topo_wu2016, LaSb_ARPES_zeng2016, LaBi_topo_ncomms13942, LaSb_topo_transition_HSE_guo2017} for potential spintronic technologies.
  
\section{CONCLUSION}
We have studied systematically rock-salt structured LaP and LaAs using first-principles calculations based on density functional theory. The employed metal-GGA functional with scissor correction method have enabled computationally efficient and precise evaluations of band structures and thermoelectric properties. We have shown that applying strain can effectively manipulate the thermoelectric performance. At the optimal carrier concentration $n=3\times 10^{20}~\textrm{cm}^{-3}$, $+2\%$ tensile strain can enhance the figure of merit $ZT$ of LaP at $T=1200$ K to $> 2$, which is 90\% larger than the unstrained value. The transport coefficients of LaAs exhibit similar carrier concentration and temperature dependences to those of LaP. 
Due to a reduced band gap and a more dispersive band structure, the thermoelectric performance of LaAs is reduced with an optimal $ZT\sim 60\%$ that of LaP.
With carrier doping and strain engineering, LaP and LaAs could be promising quantum materials for potential thermoelectric applications especially at high temperatures.

\section*{ACKNOWLEDGMENTS}
The calculations were performed on the Frontera computing system at the Texas Advanced Computing Center. Frontera is made possible by National Science Foundation award OAC-1818253.

\bibliography{bibfile_V7}

\end{document}



	\title{Supplemental Material:\\
	 First-Principles Study of Strain Effect on Thermoelectric Properties of LaP and LaAs}

	\author{Chia-Min Lin}
	\affiliation{Department of Physics, University of Alabama at Birmingham, Birmingham, Alabama 35294, USA}

	\author{Wei-Chih Chen}
    \affiliation{Department of Physics, University of Alabama at Birmingham, Birmingham, Alabama 35294, USA}
	\author{Cheng-Chien Chen}
	\affiliation{Department of Physics, University of Alabama at Birmingham, Birmingham, Alabama 35294, USA}

	\date{\today}                     

\begin{abstract}
\vspace{10mm} 
In this document, we show the computational settings and convergence tests of density functional theory (DFT) calculations for various software packages discussed in the main text.
These include structural relaxation in VASP, transport and thermoelectric properties in \textsc{BoltzTraP2}, phonon dispersion in \textsc{Phonopy}, and lattice thermal conductivity in \textsc{Phono3py}. All results presented below are based on the generalized gradient approximation (GGA) exchange-correlation energy functional.
\end{abstract}

	\pacs{Valid PACS appear here}
	\maketitle

	
	\renewcommand{\thefigure}{S\arabic{figure}}

\section{Structural Relaxation}

\begin{figure*}
	\includegraphics[width=\textwidth]{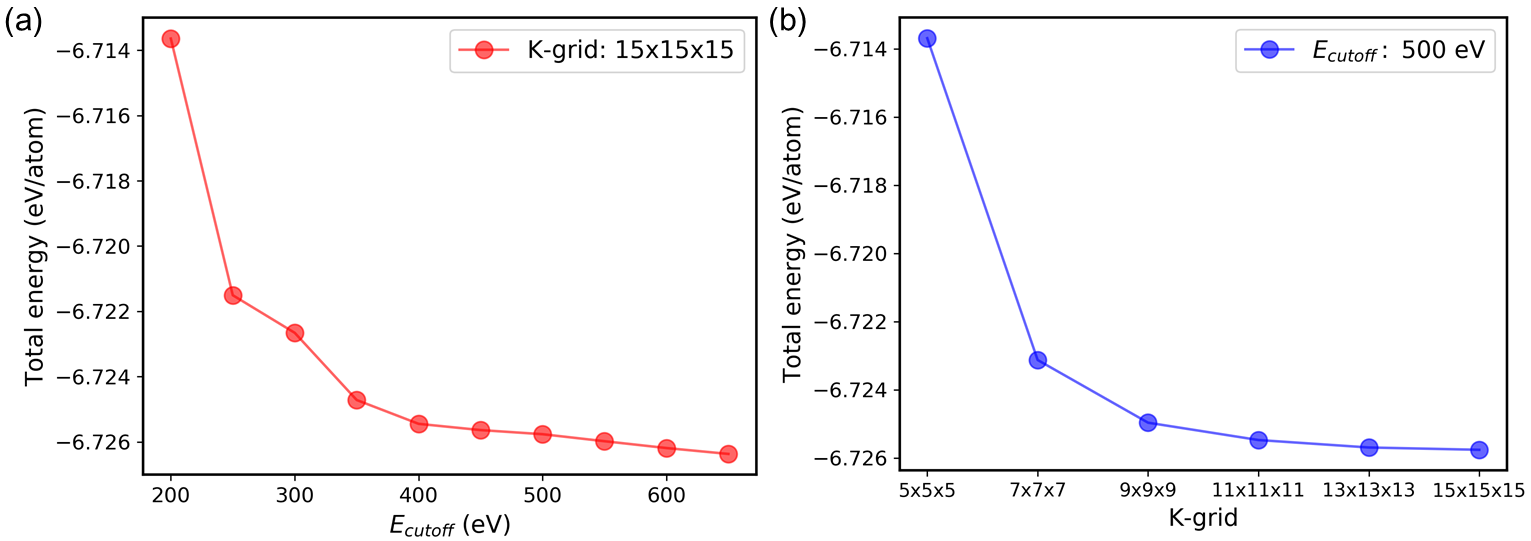}
	\caption{Convergence tests on (a) plane-wave cutoff energy, and (b) $k$-point sampling, for VASP DFT total energy of LaP.}
	\label{FIG.S1}
	\end{figure*}

In our calculations, we use the primitive cell of the rock-salt structure for the structural relaxation and thermoelectric properties. We first conduct convergence tests of plane-wave energy cutoff ($E_{\text{cutoff}}$) and $k$-grid for structural relaxation. The test results of DFT total energy for LaP are shown in FIG. \ref{FIG.S1}, which indicates that a $E_{\text{cutoff}}$ = 500 eV and a $k$-grid of 15$\times$15$\times$ 15 points is sufficient to achieve an energy convergence within $1\times10^{-4}$ eV/atom.

	\newpage
	
\section{transport and thermoelectric properties}

\begin{figure*}
	\centering
	\includegraphics[width=\textwidth]{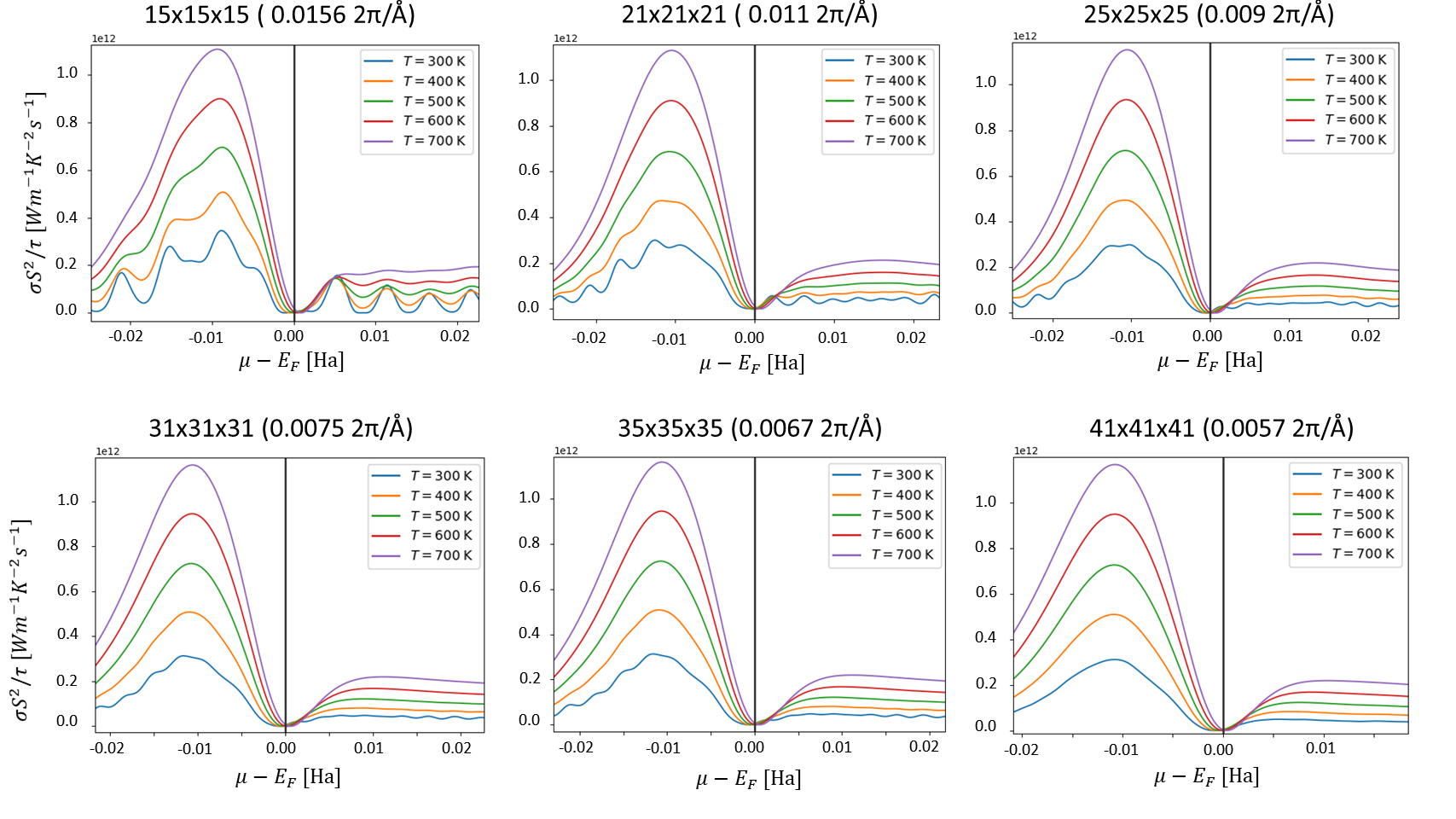}
	\caption {Convergence tests on $k$-point sampling for LaP's power factor calculations by \textsc{BoltzTraP2}. The $k$-grid size ranges from 15$\times$15 $\times$15 to 41$\times$41$\times$41 points. The result is well-converged with the 41 $\times$ 41 $\times$ 41 $k$-grid for temperature $\ge$ 300 K.}
	\label{FIG.S2}
	\end{figure*}

The calculations of transport and thermoelectric properties using \textsc{BoltzTraP2} require a very refined $k$-grid to achieve proper convergence, especially for low temperature.
Here, we show the $k$-grid convergence test on power factor ($\sigma S^2$, where $\sigma$ and $S$ are respectively the electrical conductivity and Seebeck coefficient) divided by the relaxation time $\tau$. FIG. \ref{FIG.S2} shows systematic convergence tests for $k$-grid size ranging from $15 \times 15 \times 15$ to $41 \times 41 \times 41$ points. The results show that insufficient $k$-point sampling would cause artificial oscillating behavior in the temperature-dependent power factor. The power factor is well-converged using a 41 $\times$ 41 $\times$ 41 $k$-grid for temperature $\ge$ 300 K.
	
	\newpage
	
\section{Phonon dispersion}
\begin{figure*}
	\centering
	\includegraphics[width=\textwidth]{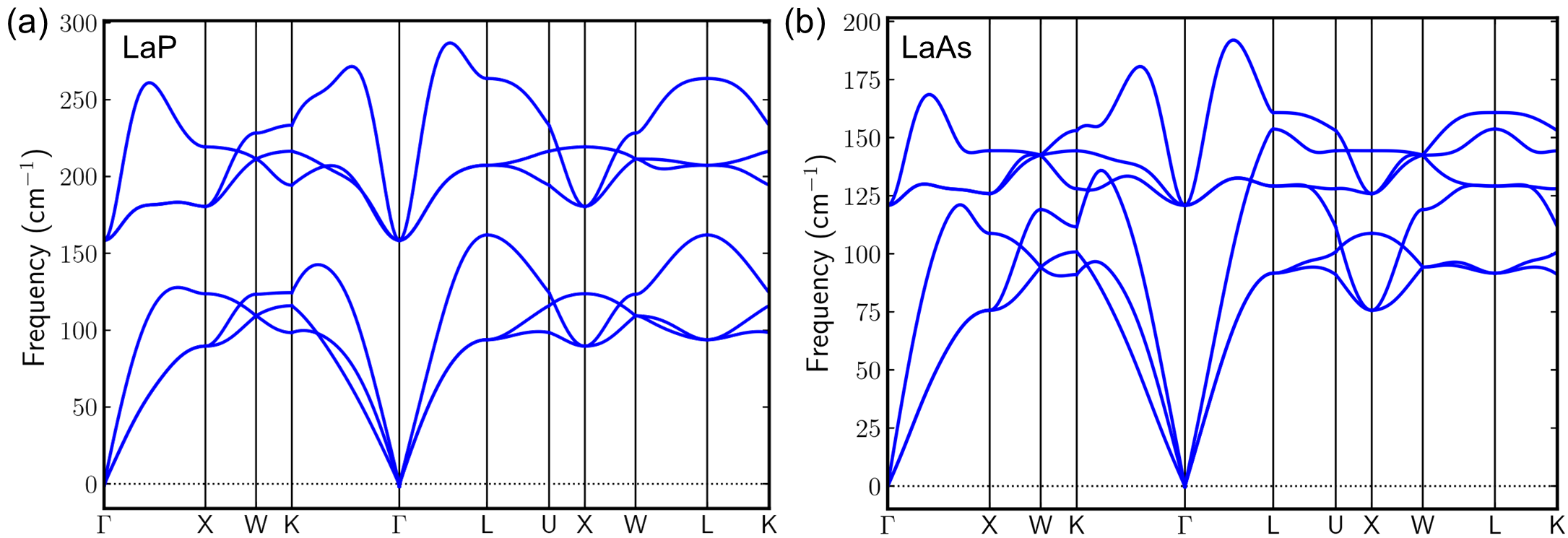}
	\caption {Phonon dispersion spectra of (a) LaP and (b) LaAs, calculated by using a 5$\times$5$\times$5 supercell with the finite-displacement method in \textsc{Phonopy}.}
	\label{FIG.S3}
	\end{figure*}

   	In FIG. \ref{FIG.S3}, we show the phonon dispersion spectra for LaP and LaAs. The calculations are performed using a 5$\times$5$\times$5 supercell with the finite-displacement method. The absence of any imaginary (negative) phonon mode indicates dynamical stabilities of these structures.
   	
	\newpage
	
\section{Energy cutoff, $q$-point, and neighbor cutoff convergence tests for lattice thermal conductivity}

\begin{figure*}
	\centering
	\includegraphics[width=\textwidth]{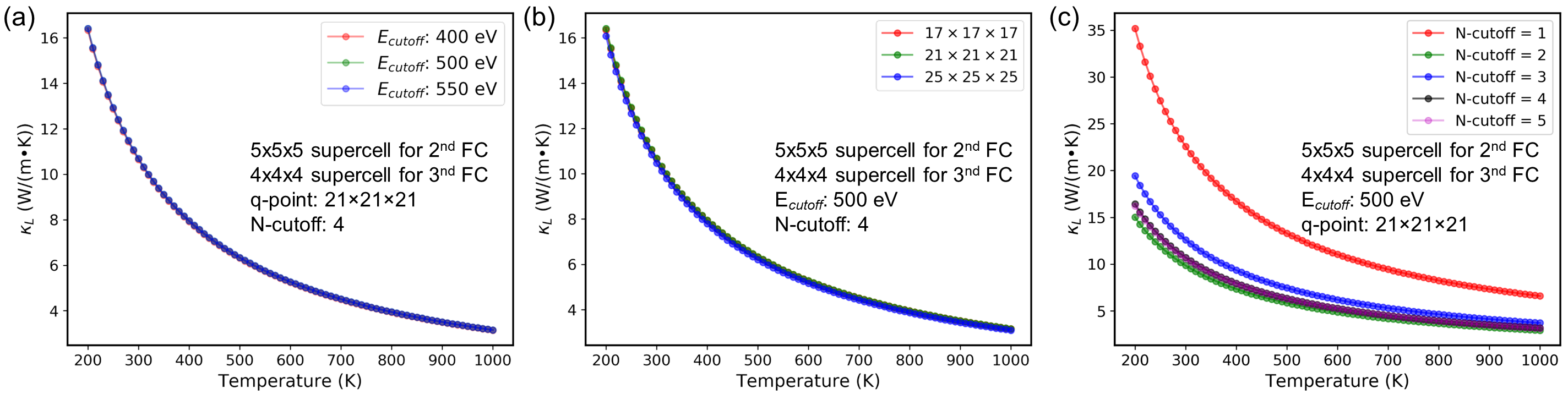}
	\caption {Convergence tests for (a) $E_\text{cutoff}$, (b) $q$-point, and (c) neighbor cutoff $N$-cutoff, on LaP lattice thermal conductivity calculated by \textsc{Phono3py}.}
	\label{FIG.S4}
    \end{figure*}

	In FIG. \ref{FIG.S4}, we perform convergence tests on energy cutoff $E_{\textrm{cutoff}}$, $q$-point, and neighbor cutoff $N$-cutoff, for the lattice thermal conductivity $\kappa_L$. The calculations utilize 5$\times$5$\times$5 and 4$\times$4$\times$4 supercells for the second- and third-order force constants, respectively. We find that the lattice thermal conductivity is converged using a 500 eV $E_\text{cutoff}$ and a 21$\times$21$\times$21 $q$-point sampling. Besides, we find that $N$-cutoff = 4 (or radius cutoff $\sim 6.1~\AA$ for unstrained LaP) is sufficient for the third-order force constants to achieve converged lattice thermal conductivity.

